\documentclass[10pt,twocolumn,twoside]{IEEEtran}
\usepackage{color}
\usepackage{cite}
\usepackage{epsfig}
\usepackage{amssymb}
\usepackage{amsmath}
\usepackage{amsthm}
\usepackage{bm}
\usepackage{booktabs}
\usepackage{multirow}
\usepackage{hyperref}

\graphicspath{{Figures/}} 

\newcommand{\Ho}{\mathcal{H}_0}
\newcommand{\Hu}{\mathcal{H}_1}
\newcommand{\Pmd}{\mathbb{P}_{\mathrm{md}}}
\newcommand{\Pfa}{\mathbb{P}_{\mathrm{fa}}}
\newcommand{\Pv}{\mathbb{P}_v}
\renewcommand{\Pi}{\mathbb{P}_\infty}
\newcommand{\Po}{\mathbb{P}_1}
\newcommand{\Ev}{\mathbb{E}_v}
\newcommand{\E}{\mathrm{E}}

\newcommand{\Ei}{\mathbb{E}_\infty}
\newcommand{\ma}{m_\alpha}
\newcommand{\Tw}{T_{\mathrm{WL}}}
\newcommand{\Ts}{T_{\mathrm{S}}}
\newcommand{\Tc}{T_{\mathrm{C}}}
\newcommand{\Tf}{T_{\mathrm{F}}}
\newcommand{\LLR}{\mathrm{LLR}}
\newcommand{\inters}{\bigcap\limits}

\newcommand{\bt}{\tilde{\beta}}
\newcommand{\at}{\tilde{\alpha}}

\newcommand{\Pf}{\mathcal{P}_\mathrm{{fa}}}
\newcommand{\tta}{t_{\mathrm{tta}}}
\newcommand{\ta}{t_\alpha}
\newcommand{\vlam}{{\boldsymbol{\lambda}}}
\newcommand{\vlamt}{\tilde{\vlam}}

\newcommand{\N}{\mathcal{N}}

\def\CN{{C/N$_0$ }}
\newcommand{\Tcn}{T_{\mathrm{F,C}}}
\newcommand{\bcn}{\beta_{\mathrm{C}}}
\newcommand{\hc}{h_{\mathrm{c}}}
\newcommand{\Xc}{x_{n,\mathrm{c}}}

\newcommand{\yc}{y}

\newcommand{\sigc}{\sigma_{\mathrm{c}}}
\newcommand{\sigcc}{\sigma_{\mathrm{c}}^2}
\newcommand{\muc}{\mu_\mathrm{c}}
\newcommand{\mtc}{\tilde{\mu}_{\mathrm{c},1}}

\newcommand{\Tdll}{T_{\mathrm{F,D}}}
\newcommand{\bdll}{\beta_{\mathrm{D}}}
\newcommand{\hd}{h_{\mathrm{d}}}
\newcommand{\Xd}{x_{n,\mathrm{d}}}

\newcommand{\sigdo}{\sigma_{\mathrm{d},0}}
\newcommand{\sigdu}{\sigma_{\mathrm{d},1}}
\newcommand{\sigddo}{\sigma^2_{\mathrm{d},0}}
\newcommand{\sigddu}{\sigma^2_{\mathrm{d},1}}

\newcommand{\std}{\tilde{\sigma}_{\mathrm{d},1}}
\newcommand{\stdd}{\tilde{\sigma}^2_{\mathrm{d},1}}

\newcommand{\Tsam}{T_{\mathrm{F,S}}}
\newcommand{\bsam}{\beta_{\mathrm{S}}}
\newcommand{\hs}{h_{\mathrm{s}}}
\newcommand{\Xs}{x_{n,\mathrm{s}}}

\newcommand{\Zs}{Z_{\mathrm{s}}}
\newcommand{\sigso}{\sigma_{\mathrm{s},0}}

\newcommand{\sigsso}{\sigma^2_{\mathrm{s},0}}
\newcommand{\sigssu}{\sigma^2_{\mathrm{s},1}}
\newcommand{\muso}{\mu_{\mathrm{s},0}}
\newcommand{\musu}{\mu_{\mathrm{s},1}}
\newcommand{\sts}{\tilde{\sigma}_{\mathrm{s},1}}
\newcommand{\stss}{\tilde{\sigma}^2_{\mathrm{s},1}}
\newcommand{\mts}{\tilde{\mu}_{\mathrm{s},1}}
\newcommand{\as}{a_\mathrm{s}}
\newcommand{\bs}{b_\mathrm{s}}
\newcommand{\cs}{c_\mathrm{s}}

\newcommand{\xs}{x_{\mathrm{s}}}
\newcommand{\ys}{y}
\newcommand{\zs}{z_{\mathrm{s}}}

\newcommand{\suim}{\sum_{i=1}^m}

\newcommand{\p}[1]{\left(#1\right)}
\newcommand{\corch}[1]{\left[#1\right]}
\newcommand{\llav}[1]{\left\{#1\right\}}
\newcommand{\eq}[1]{\begin{equation}{#1}\end{equation}}
\newcommand{\trunc}[5]{\left\{
  \begin{array}{lc}
    \Ho :& \N(#1,#2) \\
    \Hu :& \N(#3,#4)
  \end{array}#5
\right.
}
\newcommand{\hyp}[7]{\eq{#1#2\sim\trunc{#3}{#4}{#5}{#6}{#7}}}

\theoremstyle{remark}
\newtheorem{col}{Corollary}
\newtheorem{theorem}{Theorem}
\newtheorem{lemma}{Lemma}

\begin{document}
%\tableofcontents
\title{Performance Bounds for Finite Moving Average Change Detection Applied to GNSS}

\author{Daniel Egea-Roca$^{\star}$,  Gonzalo Seco-Granados$^{\star}$,  Jos\'e A. L\'opez-Salcedo$^{\star}$,  H. Vincent Poor$^{\dagger}$\\
{\small $^{\star}$ Department of Telecommunications and Systems Engineering, Universitat Aut\`{o}noma de Barcelona (UAB), Bellaterra, Spain\\
\small $^{\dagger}$ Department of Electrical Engineering, Princeton University, Princeton, NJ 08544, USA}
\thanks{This work was partly supported by the Spanish Grant TEC2014-53656-R, the European Commission iGNSSrx project, and the U.S. National Science Foundation under Grant ECCS-1549881.}}

\maketitle
\vspace{-2cm}
\begin{abstract}
Due to the widespread deployment of Global Navigation Satellite Systems (GNSSs) for critical road or urban applications, one of the major challenges to be solved is the provision of integrity to terrestrial environments, so that GNSS may be safety used in these applications. To do so, the integrity of the received GNSS signal must be analyzed in order to detect some local effect disturbing the received signal. This is desirable because the presence of some local effect may cause large position errors, and hence compromise the signal integrity. Moreover, the detection of such disturbing effects must be done before some pre-established delay. This kind of detection lies within the field of transient change detection. In this work, a finite moving average stopping time is proposed in order to approach the signal integrity problem with a transient change detection framework. The statistical performance of this stopping time is investigated and compared, in the context of multipath detection, to other different methods available in the literature. Numerical results are presented in order to assess their performance.
\end{abstract}


%----------------------------------------------------------------%
%-------------------------- SECTION #1 --------------------------%
%----------------------------------------------------------------%
\section{Introduction}
%% PVT Integrity Introduction %%
%------------ GNSS Integrity motivation (safety/critical apps) -----------%
Recently, there has been an increasing interest in Global Navigation Satellite System (GNSS)-based safety- and liability-critical applications \cite{salcedo}. These applications, which are often associated with terrestrial environments, have very stringent requirements in terms of accuracy, continuity and integrity of the provided position solution. Besides providing an accurate navigation solution, timely warnings must be provided to the user when the system should not be used; this capability is referred to as the \emph{integrity} of the system. 

%------------ Current GNSS Integrity (SBAS/RAIM) -----------%
The concept of GNSS integrity was originally developed for civil aviation \cite{SBAS}. In particular, civil aviation standards specify a set of minimum performance requirements for each operation and phase of flight \cite{RTCA}. Nevertheless, a standalone GNSS receiver cannot meet the stringent civil aviation requirements, and then various augmentation systems have been developed to fulfill these requirements. These systems are classified according to their infrastructure into Ground-, Satellite-, and Aircraft-Based Augmentation Systems (GBAS, SBAS and ABAS) \cite{SBAS}. They constitute powerful methods providing the user with integrity information. However, their infrastructure is very complex and therefore costly.

It is for this reason that Receiver Autonomous Integrity Monitoring (RAIM) techniques were designed based on calculations performed within the user equipment itself \cite{RAIM1,RAIM2,integrity}. These techniques rely on redundant measurements coming from different satellites, which are feasible to obtain in civil aviation. However, this is not the case in terrestrial environments, where a plethora of obstacles blocking the signal of different satellites may be present. Moreover, both SBAS and RAIM assume that local effects like multipath, interference and spoofing have a controlled influence on the GNSS signal. This is really the case for civil aviation but it is not for terrestrial environments, so that position errors and, thus, integrity are strongly affected \cite{GMV}.

%------------ Signal level integrity (sequential change detection problem) -----------%
Hence, in terrestrial-critical applications, it is of paramount importance to promptly detect any possible anomaly or misleading behavior that could be endangering the received GNSS signal. Otherwise, the safety and trust of the end-user position and time could be jeopardized. We refer to this capability as \emph{signal integrity}, which is currently a concern within the GNSS community \cite{GMV,GMV2}. Moreover, for integrity purposes, an acceptable detection delay is limited by a given value $m$, so that late detections (i.e. delayed more than $m$) are considered as missed. This kind of detection lies on the field of sequential change detection, including \emph{quickest} change detection (QCD) and \emph{transient} change detection (TCD).

%% Sequential Change Detection Introduction %% 
%------------ Quickest Change Detection -----------%
The traditional QCD problem deals with a change (i.e. anomaly) of infinite duration. The optimal criterion in this case is to minimize the detection delay subject to a level of false alarms. Comprehensive overviews of this kind of detection can be found in \cite{Poor} and \cite{Baseville}. In the present work, we focus on non-Bayesian approaches in which the change time $v$ is modeled as being unknown but non-random. For this kind of approach, the CUSUM algorithm was first proposed as a continuous inspection scheme in the 1950s \cite{Page}, but it was not until 1971 that its optimality for the QCD problem was established asymptotically (i.e. when false alarms go to zero) \cite{Lorden}. More than a decade later, Moustakides \cite{Moustakides1} proved that the CUSUM is also non-asymptotically optimal.

%------------ Transient Change Detection -----------%
In contrast, the TCD problem deals with finite change duration and then a bounded detection delay is desired. Unfortunately, the traditional QCD criterion (i.e. minimizing the detection delay) does not completely fit into this problem. In this case, we wish to minimize the probability of missed detection (i.e. late detections) subject to a level of false alarms. This criterion was first adopted in \cite{Bojdecki} for the Bayesian approach, but without controlling the false alarm rate. Recently, the authors of \cite{Pollak} considered a semi-Bayesian approach imposing a suitable constraint on the false alarm rate. Very recently, the first optimal results for the non-Bayesian case were provided by Moustakides in \cite{Moustakides2}. Nevertheless, it is worth mentioning that all of these works  have considered the very particular case of $m=1$, which has limited practical application.

\subsection{Related Literature}
In order to improve integrity in terrestrial environments we have to detect local effects as soon as possible in order to quickly alert the user. These effects are well-known problems within the GNSS community, leading to a plethora of contributions in the existing literature for interference \cite{interf1,interf2}, multipath \cite{MP1,MP2} and spoofing \cite{spoof1,spoof2} detection. However, all the existing contributions have adopted a classical detection framework, which is not well suited to fulfill the integrity requirements of safety-critical applications. For the prompt detection of integrity threats, it is essential to formulate the problem under the framework of sequential change detection.

The QCD framework has been extensively applied to a wide range of fields such as distributed detection in sensor networks \cite{WSN}, signal detection in multi-antenna receivers \cite{signal_detection} and spectrum sensing in cognitive radio \cite{cognitive}, just to mention a few. Nevertheless, it has barely been used in the GNSS arena. Based on this observation, we already addressed the problem of detecting local degrading effects in a QCD framework. Specifically, we addressed the problem for the case of multi-antenna GNSS receivers (see \cite{SAM,ICL14}), and for single-antenna receivers in \cite{ICL15} and \cite{VTC}. Finally, a comprehensive overview of interference and multipath quickest detection was provided in \cite{IONint} and \cite{ION}, respectively, including an extensive analysis with real signals. The results in \cite{ION} corroborated the improvements of incorporating signal integrity into the positioning accuracy and integrity.

Nonetheless, all of these contributions adopted the framework of QCD. However, for integrity purposes, a bounded delay is desirable and then we should rely on the framework of TCD. Moreover, we are interested in non-Bayesian approaches. Several approaches have been proposed in the last decade. Standard solutions are based on the CUSUM algorithm \cite{Willett,Willett1,Willett2,Nikiforov,Guepie}. Unfortunately, almost all available results are applicable to off-line detection on finite observation intervals and they adopt the traditional criterion of QCD. Exceptions, adopting the probability minimization criterion, are \cite{Nikiforov}, which applies the CUSUM algorithm for GNSS integrity (at position level), and \cite{Guepie}, which proposes a windowed version of the CUSUM for monitoring the quality of drinking water.

\subsection{Contributions}
%% Contribution %%
The TCD framework has not been applied so far to the problem of signal integrity. Moreover, for the more general case of finite $m>1$, which is the case of GNSS signal integrity, and to the best of the authors' knowledge, no optimal solution has been found for the TCD problem. On the other hand, the Finite Moving Average (FMA) stopping time has been shown to be the optimal windowed CUSUM solution (i.e. using the last $m$ samples) for the Gaussian mean change case \cite{Guepie}. Based on these observations, we propose the use of an FMA stopping time for approaching the general problem of TCD, in general, and signal integrity monitoring in GNSS, in particular.

Hence, the contribution of this work is twofold. Firstly, we theoretically investigate the statistical performance of the FMA stopping time, based on the new optimal criterion of probability minimization, for any general case. This leads to the provision of novel bounds valid for any kind of change and not restricted to the Gaussian mean change. These bounds are more tight than those available in the literature for other approaches, drastically benefiting the availability of the integrity algorithm. These bounds were briefly introduced in \cite{SSP}. In this work, though, we provide a more extensive and complete proof. 
%Secondly, we provide a rigorous formulation of the signal integrity problem in GNSS. This formulation fills the gap between local threat detection and TCD. The application of the FMA stopping time is suggested for approaching the signal integrity problem with a TCD framework. This is a new approach and it is intended to provide the level of integrity needed for terrestrial environments, which currently cannot be achieved by means of classical integrity techniques. 

Finally, we show with numerical results that the FMA stopping time outperforms other methods available in the literature for the TCD. This is done in the setting of signal level integrity in GNSS, when dealing with multipath detection. This was briefly introduced in \cite{SSP} for the carrier-to-noise ratio (C/N$_0$) metric, which was modeled as a Gaussian mean change. Here, we provide a more extensive analysis including the three multipath detection techniques presented in \cite{ION}, including all possible changes in a Gaussian distribution (i.e. mean or/and variance changes). Moreover, we present numerical results analyzing the availability of the proposed signal integrity algorithm.

The rest of the paper is organized as follows: Section \ref{sec:back} provides background on sequential change detection, including both QCD and TCD, and introducing the proposed FMA stopping time. Next, Section \ref{sec:stat} presents the statistical properties of the FMA stopping time. Then, Section \ref{sec:metrics} investigates the application of these results to the analyzed multipath detection metrics. Finally, Section \ref{sec:results} presents our numerical results, while Section \ref{sec:conclusions} concludes the paper.

%------------------------------------------------------------------------%
%-------------------------- SECTION 2 -----------------------------------%
%------------------------------------------------------------------------%
\section{Background on sequential change detection}
\label{sec:back}
A change detection algorithm, including QCD and TCD, is completely defined by its stopping time $T$ at which the change is declared. In general, a change detection algorithm can be modeled as follows: Let $\{x_n\}_{n\geq1}$ be a random sequence observed sequentially providing information about some integrity threat, and let $v$ be the instant (in samples) when the integrity threat appears. We consider a family $\{\Pv|v\in [1,2,\dots,\infty]\}$ of probability measures, such that, under $\Pv$, $x_1,\dots,x_{v-1}$ and $x_{v+m},\dots,x_{\infty}$, with $m$ the change duration, are independent and identically distributed (iid) with a fixed marginal probability density function (pdf) $f_0$, corresponding to the normal conditions (i.e. the integrity threat is not present, $\Ho$). On the other hand, $x_{v},\dots,x_{v+m-1}$ are iid with another marginal pdf $f_1 \neq f_0$, corresponding to the abnormal conditions (i.e. the integrity threat is present, $\Hu$).

In this section, we firstly briefly recall the problem of QCD. Secondly, we introduce the problem of TCD. Finally, the idea of windowed solutions and the FMA stopping time proposed in this paper are presented.

\subsection{Quickest change detection (QCD): CUSUM stopping time}
The statistical model for QCD is described as follows:
\eq{\label{eq:QDmodel}
x_n\sim
\left\{
  \begin{array}{lcl}
    \Ho :& f_0(x)& \mbox{if } n<v\\
    \Hu :& f_1(x)& \mbox{if } n\geq v
  \end{array},
\right.
}
where in this case $m=\infty$.

Denoting $\Ev$ as the expectation under the probability measure $\Pv$ the effectiveness of QCD has been traditionally quantified by the minimization of \cite{Lorden}:
\eq{\label{eq:QDcrit}
d(T) = \sup_{v\geq1}\text{essup }\Ev\corch{\p{T-v+1}^+|x_1,\dots,x_{v-1}}
}
among all stopping times $T$ satisfying $\Ei(T)\geq\gamma$, where $(x)^+ = \max(0,x)$, $\text{essup}$ denotes the essential supremum, and $\gamma>0$ a finite constant. That is, we seek a stopping time $T$ that minimizes the delay $d(T)$ within a lower-bound constraint on the mean time between false alarms $\Ei(T)$.

The following CUSUM stopping time was proposed in \cite{Page}:
\eq{
\Tc(h) \doteq \inf\llav{n\geq 1: \max_{1\leq k\leq n}S_k^n\geq h}; S_k^n \doteq \sum_{i=k}^n \LLR(i),
}
where $\LLR(i) \doteq \ln(f_1(x_i)/f_0(x_i))$ is the log-likelihood ratio (LLR) of the observation $x_i$ and $h$ is the detection threshold. The first optimal results of the CUSUM in the sense of the criterion in \eqref{eq:QDcrit} were shown in \cite{Lorden} and \cite{Moustakides1}, in an asymptotic (i.e. $\gamma \to \infty$) and non-asymptotic (i.e. for all finite $\gamma$) way, respectively. However, as shown in \cite{Lai}, the requirement of having large values of $\Ei(T)$ does not guarantee small values of the probability of false alarm $\Pi(l\leq T < l+m_\alpha)$ within a fixed interval of length $\ma$, for all $l\geq 1$. As a result, \cite{Lai} proposed to replace the traditional constraint $\Ei(T)\geq\gamma$ by the following constraint on the worst-case probability of false alarm within any interval of length $\ma$, which is more convenient for safety-critical applications:
\eq{\label{eq:Pfac}
\Pfa\p{T,\ma} \doteq \sup_{l\geq 1} \Pi\p{l\leq T < l+m_\alpha}\leq \alpha.
}
It was shown in \cite{Lai} that the CUSUM stopping time $\Tc$ asymptotically minimizes (as $\Pfa\to 0$) the detection delay, over all stopping times $T$ satisfying \eqref{eq:Pfac}, if $h$ fulfills the following equation:
\eq{\label{eq:Tcbound}
\Pfa\p{\Tc,\ma}\leq\ma e^{-h}.
}

\subsection{Transient change detection (TCD): Shewhart stopping time}
Unlike QCD, in which the change duration $m$ is assumed to be infinite, the change duration in TCD problems is assumed to be finite. This is modeled as
\eq{\label{eq:transientmodel}
x_n \sim 
\left\{
  \begin{array}{lcl}
    \Ho :& f_0(x)& \mbox{if } n < v \mbox{ or } n \geq v+m\\
    \Hu :& f_1(x)& \mbox{if } v\leq n < v+m
  \end{array}.
\right.
}
As discussed in \cite{Guepie}, there are two types of TCD problems. The first type involves the detection of suddenly arriving signals of random unknown duration. In this case $m$ denotes the unknown duration of the change. The second type involves safety-critical applications where the maximum permitted detection delay is a priori fixed to a pre-established value, and then $m$ is considered to be known. 

We observe from \eqref{eq:QDcrit} that no hard limit is imposed on the detection delay; consequently, this quantity can become arbitrarily large. In this sense, the optimality criteria for the TCD problem should be modified in order to seek a small probability of missed detection given an acceptable false alarm rate. In other words, we wish to have $v \leq T < v+m$. Stopping within the prescribed interval constitutes a desirable event while stopping at $T\geq v+m$ is considered a missed detection. This is equivalent to the optimality criterion introduced in \cite{Moustakides2} and \cite{Guepie}, used through this paper, which involves the following minimization:
\eq{\label{eq:transient1}
\inf_{T\in C_\alpha}\llav{\Pmd(T,m_\mathrm{d}) \doteq \sup_{v\geq1}\Pv\p{T\geq v+m_\mathrm{d}|T\geq v}}
}
among all stopping times $T\in C_\alpha$ satisfying
\eq{\label{eq:transient2}
C_\alpha = \llav{T: \Pfa(T,\ma) \leq \alpha},
}
where $m_\mathrm{d}$ denotes the maximum permitted detection delay and $\Pmd$ and $\Pfa$ stand for the worst-case probabilities of missed detection and false alarm within any interval of length $\ma$, respectively, with $\Pfa$ defined as in \eqref{eq:Pfac}. As we will see, this criterion is very convenient for GNSS integrity monitoring.

This article belongs to the second class of TCD problem stated above (i.e. safety-critical applications); in this case a detection with a delay greater than $m_\mathrm{d}$ is considered as missed, even if $m>m_\mathrm{d}$. On the other hand, if the duration of the change $m$ is smaller than $m_\mathrm{d}$, then such a change is considered less dangerous because its impact on the system is limited or negligible. It is for this reason that the duration $m$ is considered to be known henceforth and equal to $m_\mathrm{d}$ (i.e. $m=m_\mathrm{d}$). Indeed, for integrity algorithms, we can fix $m$ as $m=F_{\mathrm{s}}\cdot\tta$ (similarly $\ma=F_{\mathrm{s}}\cdot\ta$), with $F_{\mathrm{s}}$ the sampling rate of $x_n$, and $\tta$ the so-called time to alert (TTA), which is given by norms and standards \cite{RTCA}, as well as $\ta$. Very recently, the optimal solution of this class of TCD problem, for $m=1$, was shown to be the Shewhart test \cite{Moustakides2}
\eq{
\Ts(h) \doteq \inf\llav{n\geq 1: \LLR(n)\geq h}.
} 
Indeed, this is the only available result on optimality for the criterion in \eqref{eq:transient1}--\eqref{eq:transient2}. Nevertheless, for the case of finite $m>1$, which is the case of integrity monitoring, no optimal solution has been found yet, so the problem is still open.

\subsection{Windowed solutions: WLC and FMA stopping times}
Since there is no optimal solution available in the literature of TCD for a finite $m>1$, we propose a windowed solution based on the following idea: We know that the optimal solution for QCD, that is for $m = \infty$, is the CUSUM test \cite{Moustakides1}, which uses information about all the past samples. On the other hand, the Shewhart test, which uses information of one sample, is established to be optimal for the non-Bayesian TCD problem with $m=1$ \cite{Moustakides2}. Hence, it is intuitive to think that the optimal solution for $1<m<\infty$ would be some test statistic between these two techniques, and particularly, a test statistic using information about $m$ samples (i.e. windowed). 

In this context, a window-limited CUSUM (WLC) test is proposed in \cite{Guepie} by using at each moment the $m$ last observations, only. The stopping time is given by
\eq{
\Tw\p{h} \doteq \inf\llav{n\geq m: \max_{n-m+1\leq k\leq n} S_k^n\geq h}.
}
It is assumed that the WLC is not operational during the first $m-1$ observations. It is also shown that the optimal WLC stopping time, with respect to the criterion in \eqref{eq:transient1}-\eqref{eq:transient2}, for a Gaussian mean change leads to
\eq{
\Tw^*\p{\tilde{h}} = \inf\llav{n\geq m: \sum_{i=n-m+1}^n x_i \geq \tilde{h}},
}
where $\tilde{h}$ denotes the chosen threshold.

Inspired by this result and the idea of windowed solution, in this paper we propose the use of an FMA stopping time for any general TCD problem, which becomes
\eq{\label{eq:Tfma}
\Tf\p{h} \doteq \inf\llav{n\geq m: S_n\geq h} ; S_n = S_{n-m+1}^n.
}
It is worth noting that this stopping time is equivalent to an FMA test with the LLR of a Gaussian mean change. Furthermore, the use of the FMA stopping time is motivated by the fact, as we will show next, that we can obtain tight bounds for both the probabilities of missed detection and false alarm. On the contrary, the bounds available in the literature for these probabilities for the CUSUM and WLC are not that tight. This is a key factor for the integrity problem since it considerably improves the availability of the algorithm.

%------------------------------------------------------------------------%
%--------------------- SECTION 3 ----------------------------------------%
%------------------------------------------------------------------------%
\section{Statistical Performance of the FMA Stopping Time}
\label{sec:stat}
The goal of this section is to theoretically investigate the statistical performance of the FMA stopping time $\Tf(h)$; that is, to determine the worst-case probability of missed detection $\Pmd\p{\Tf(h),m}$ and the worst-case probability of false alarm for a given duration $\ma$, $\Pfa\p{\Tf(h),\ma}$. The exact calculation of these probabilities is very complicated, and it is here where the existence of tight-enough bounds is of practical interest. These bounds are stated in the next theorem.
\begin{theorem}\label{th:stat}
Let us consider the FMA stopping time $\Tf(h)$ in \eqref{eq:Tfma} and let $S_m = S_1^m = \sum_{i=1}^m \LLR(i)$; then the worst-case probability of false alarm for a given duration $\ma$ is bounded as
\eq{\label{eq:Pfaineq}
\Pfa\p{\Tf(h),\ma} \leq \alpha\p{h,\ma},
}
where
\eq{\label{eq:alpha}
\alpha\p{h,\ma} = 1 - \corch{\Pi\p{S_m < h}}^{\ma}.
}
On the other hand, the worst-case probability of missed detection is bounded as
\eq{\label{eq:Pmdineq}
\Pmd\p{\Tf(h),m} \leq \beta(h,m),
} 
where 
\eq{\label{eq:beta}
\beta(h,m) = \mathbb{P}_1\p{S_m < h}.
}
\end{theorem}
\begin{IEEEproof}
The proof is given in Appendix \ref{app:stat}.
\end{IEEEproof}

In practice, values $\alpha$ for $\Pfa$ are imposed, so that we have to guarantee that $\Pfa \leq \at$. Thus, the threshold $h$ has to be selected in order to satisfy this constraint, and then $\Pmd$ turns out to be a function of the fixed $\alpha$ (i.e. $\Pmd\p{\Tf(h(\at)),m}$). In some sense, $\Pmd\p{\Tf(h(\at)),m}$ plays the same role in the TCD theory as the Cram\'er-Rao lower bound in estimation theory, or as the receiver operating characteristic (ROC) in classical detection. Moreover, this kind of ROC allow us to compare the performance of different algorithms in terms of the optimality criterion in \eqref{eq:transient1}. This relation between $\Pmd$ and $\Pfa$ is given in the following corollary.
\begin{col}\label{col:stat}
Let $F_i$, with $i=\{0,1\}$, be the cumulative distribution function (cdf) of $S_m$ under $\mathcal{H}_i$ and let $h$ be selected so that
\eq{
\Pfa(\Tf(h),\ma) \leq \at,
}
with $\at$ a desirable constant value for the probability of false alarm. Hence
\eq{\label{eq:h}
h\p{\at} = F_0^{-1}\corch{\p{1-\at}^{1/\ma}},
}
where $F_0^{-1}$ is the inverse of $F_0$, and thus
\eq{\label{eq:betah}
\beta\p{h\p{\at},m} = F_1\corch{F_0^{-1}\corch{\p{1-\at}^{1/\ma}}}.
}
Moreover,
\eq{\label{eq:Pmdh}
\Pmd\p{\Tf\p{\tilde{h}},m} \leq \beta\p{h\p{\at},m},
}
with $\tilde{h}$ the threshold such that $\Pfa\p{\Tf\p{\tilde{h}},\ma}=\at$.
\end{col}
\begin{IEEEproof}
It is straightforward to see that $\mathbb{P}_j(S_m<h) = F_i(h)$, with $j=\{\infty,1\}$ and $i=\{0,1\}$, is the cdf of $S_m$ under $\mathcal{H}_i$, respectively, evaluated at $h$. Hence, from \eqref{eq:Pfaineq} and \eqref{eq:alpha} we can fix the threshold $h$ as
\eq{\label{eq:56}
h = F_0^{-1}\corch{\p{1-\at}^{1/\ma}},
}
guaranteeing that $\Pfa\p{\Tf(h),\ma}\leq \at$, and \eqref{eq:h} thus follows. The proof of \eqref{eq:betah} follows immediately by the definition of the cdf $F_1$ and by substituting \eqref{eq:56} into \eqref{eq:beta}. In order to prove \eqref{eq:Pmdh} it is worth noting that the threshold $\tilde{h}$, such that $\Pfa\p{\Tf\p{\tilde{h}},\ma}=\at$, is lower than $h$ (i.e. $\tilde{h}<h$) and that
\eq{
\Pmd\p{\Tf\p{\tilde{h}},m} \leq \beta\p{\tilde{h},m} \leq \beta\p{h,m},
}
where the last inequality follows because $\beta(h,m)$ in \eqref{eq:beta} is a monotonically increasing function on $h$, so that \eqref{eq:Pmdh} follows.
%Finally, \eqref{eq:Pmdh} follows from \eqref{eq:beta}.
\end{IEEEproof}

The previous results are valid for the general FMA stopping time. That is, they are not restricted to the Gaussian mean change, as derived in \cite{Guepie}, but they are valid for any kind of change. Moreover, as we will see later, these bounds are tighter than bounds for other available methods in the literature. This is beneficial for the availability of integrity algorithms. Unfortunately, we cannot establish the optimality of the proposed FMA stopping time in the class $C_\alpha$. To do so we should analyze the speed of convergence of the term $F_0^{-1}[(1-\alpha)^{1/\ma}]$, which is beyond the scope of this paper. Nonetheless, in \cite{Guepie} the optimality of the FMA stopping time is shown for the particular case of a Gaussian mean change. In addition, we will show later how the proposed FMA stopping time outperforms other available methods in the literature. This makes it evident that the FMA stopping time is a good candidate for TCD problems, and thus for integrity monitoring. 
%For the sig-RAIM, we fix the detection threshold so that $\Pfa\leq \at$ as in \eqref{eq:h} and we bound the integrity risk as in \eqref{eq:betah} (i.e. $\Pmd = \beta(h,m)$. Hence, if $\Pmd \leq \bt$ the sig-RAIM is declared available.

%------------------------------------------------------------------------%
%--------------------- SECTION 4 ----------------------------------------%
%------------------------------------------------------------------------%
\section{Signal Integrity Metrics}
\label{sec:metrics}
Integrity algorithms are based on the so-called integrity risk, which is defined as the probability that an integrity threat is present without raising an alarm within the TTA. This is equivalent to the probability of missed detection defined in \eqref{eq:transient1}. On the other hand, constraints on the false alarms are given in the form of \eqref{eq:transient2}. It is for this reason that the TCD framework perfectly fits with integrity algorithms. Specifically, values $\bt$ and $\at$ for the integrity risk and probability of false alarms are given by norms and standards. Then, the detection threshold is fixed so that $\Pfa\leq \at$ and if $\Pmd \leq \bt$ the algorithm is declared available, otherwise the algorithm is not available because it cannot guarantee integrity within the desired requirements. If the algorithm is available, an alarm will be raised based on some integrity metric able to detect integrity threats. For details on GNSS integrity algorithms see \cite{SBAS}.

We focus on signal integrity, henceforth termed sig-RAIM. One of the crucial factors in sig-RAIM is the computation of the signal integrity metric. Indeed, we will have different metrics providing information about different signal integrity threats. These threats come from the local effects we may encounter in terrestrial environments, mainly including multipath, radio-frequency interference and spoofing. Hence, different metrics should be available for detecting these kinds of local effects. Moreover, for the same threat (e.g. interference) we should use different metrics in order to be able to detect a wide range of situations covering different behaviors of the same threat (e.g. wide-band, narrow-band, \dots) \cite{IONint,ION}. The key point of these metrics is that they are computed from signal features such as the statistical properties, time-frequency behavior, correlation function, etc., and so we do not need additional measurements coming from different satellites (as is the case of traditional RAIM \cite{integrity}). This is important because it allows computing these metrics in terrestrial environments, where redundant information is scarce. 

A complete overview of interference and spoofing detection metrics can be found in \cite{IONint} and \cite{spoof1}, respectively. In this work, we focus on multipath detection metrics, referred as to signal integrity metrics henceforth. In particular, the presented metrics in this work are introduced in \cite{ION} in a QCD framework, including the C/N$_0$, the code discriminator output (DLLout) and the slope asymmetry metric (SAM), which are measurable metrics after the correlation process in a GNSS receiver. In order to fix the detection threshold $h$ and to compute the integrity risk $\Pmd$ given by this threshold (rather $\beta(h,m)$), which are needed for the operation of sig-RAIM, we have to obtain first the cdf of $S_m = \sum_{i=1}^m \LLR(i)$, which depends on the signal integrity metric used to feed the detection algorithm. 

In general, the presented signal integrity metrics can be modeled as Gaussian random variables \cite{ION}, and then we model the problem as a change on a Gaussian distribution:
\eq{
x_n\sim
\left\{
  \begin{array}{lcl}
    \Ho :& \N(\mu_0,\sigma_0^2)& \mbox{if } n < v \mbox{ or } n\geq v+m \\
    \Hu :& \N(\mu_1,\sigma_1^2)& \mbox{if } v\leq n <v+m
  \end{array},
\right.
}
where $\N(\mu_i,\sigma_i^2)\doteq (1/\sqrt{2\pi\sigma_i^2})\exp(-(x-\mu_i)^2/2\sigma_i^2)$ is the normal pdf with mean $\mu_i$ and variance $\sigma_i^2$, given by the mean and variance of $x_n$ under $\mathcal{H}_i$. Hence, in general, we have
\eq{\label{eq:LLRgen}
\LLR(n) = \ln\p{\frac{\sigma_0}{\sigma_1}} + \frac{\p{x_n - \mu_0}^2}{2\sigma_0^2} - \frac{\p{x_n - \mu_1}^2}{2\sigma_1^2}.
}
It is worth pointing out that the parameters under $\Hu$ are unknown, and so we will use tuning parameters ($\vlamt$) with the aim of avoiding nuisance parameters. These parameters should be fixed as the minimum change we want to detect, which will be given by
\eq{\label{eq:PL}
\vlamt \doteq q(\epsilon),
}
with $\epsilon$ the maximum allowable positioning error without issuing an alert and $q(\cdot)$ mapping positioning errors to change parameters.

The goal of this section is to statistically characterize the multipath detection metrics to use them for sig-RAIM. This is done by using the general expression of the LLR in \eqref{eq:LLRgen} and particularizing it for the different metrics presented in this work. So, we obtain the LLR distribution for the different metrics and then we obtain the cdf of $S_m$. We start with the simpler cases of having a change in either the mean or variance of a Gaussian distribution, which are the cases of the \CN and DLLout metrics, respectively. Finally, we present the most general case of having both a change in the mean and variance of a Gaussian distribution, which is the case of the SAM metric. 

\subsection{\CN metric (Gaussian mean change)}
In this section, we analyze the statistical characterization of the LLR for the C/N$_0$ metric. First, let us introduce the statistical model, given in \cite{ION}
\hyp{\label{eq:CN0model}}{\Xc}{\muc{_{,0}}}{\sigcc}{\muc{_{,1}}}{\sigcc}{,}
with $\Xc$ the \CN samples, $\muc{_{,0}}$ and $\sigcc$ the known mean and variance of $\Xc$ under $\Ho$, and $\muc{_{,1}}$ the unknown change parameter, denoting the mean of $\Xc$ under $\Hu$. For the sake of notational simplicity, we have omitted the time conditions in each hypothesis (i.e. if $n<v$ or $n>v+m$, \dots) corresponding to the TCD model. Thereby, we model the appearance of multipath as a change in the mean of a Gaussian distribution, and then the following result is obtained.
\begin{col}\label{col:CN0}
Let $\mtc$ and $\muc{_{,1}}$ be the tuned change parameter, as in \eqref{eq:PL}, and the actual change parameter, respectively, and let $\muc{_{,0}}$ and $\sigc$ be known. Therefore the LLR for the \CN is given by
\eq{\label{eq:LLRc}
\LLR_\mathrm{c}(n)= y_n= \frac{\mtc - \muc{_{,0}}}{\sigcc}\p{\Xc - \frac{\mtc + \muc{_{,0}}}{2}},}
with mean and variance equal to
\eq{\label{eq:muvarc}
\begin{aligned}
\mu_{\yc,0} &= -\frac{\p{\mtc - \muc{_{,0}}}^2}{2\sigcc}, \quad \sigma^2_{\yc} = -2\mu_{\yc,0},  \\
\mu_{\yc,1} &= \frac{\mtc - \muc{_{,0}}}{\sigcc}\p{\muc{_{,1}} - \frac{\mtc + \muc{_{,0}}}{2}}.
\end{aligned}
}
Now, let $\Tcn(\hc)$ be the FMA stopping time in \eqref{eq:Tfma} for the \CN metric with LLR in \eqref{eq:LLRc} and threshold $\hc$ so that $\Pfa(\Tcn(\hc),\ma)\leq \at$, and let $\Phi(x)$ denote the cdf of the standard normal distribution. Hence
\begin{align}
\label{eq:colc1}
&h_\mathrm{c}\p{\at} = \sqrt{m\sigma^2_{\yc}}\Phi^{-1}\corch{\p{1-\tilde{\alpha}}^{1/\ma}} +  m\mu_{\yc,0},\\
\label{eq:colc2}
&\Pmd(\Tcn\p{h_\mathrm{c}},m) \leq \bcn(\hc,m) = \Phi\p{\frac{h_\mathrm{c} - m\mu_{\yc,1}}{\sqrt{m\sigma^2_{\yc}}}}, \\
\label{eq:colc3}
&\Pfa(\Tcn\p{\hc},\ma)  \leq  1 - \corch{\Phi\p{\frac{h_\mathrm{c} - m\mu_{\yc,0}}{\sqrt{m\sigma^2_{\yc}}}}}^{\ma}.
\end{align} 
\end{col}
\begin{IEEEproof} 
The change in the \CN metric is modeled as a Gaussian mean change and thus, substituting $\mu_0 = \muc{_{,0}}$, $\mu_1 = \mtc$ and $\sigma_0 = \sigma_1 = \sigc$ into \eqref{eq:LLRgen}, \eqref{eq:LLRc} follows after simple calculus. Thereby, it is trivial to see, from \eqref{eq:LLRc} and \eqref{eq:CN0model}, that the LLR for the \CN is Gaussian distributed in both hypotheses, with mean and variance as in \eqref{eq:muvarc}. Hence, $S_m^{(\mathrm{c})}=\sum_{i=1}^m \LLR_\mathrm{c}(i)$ is Gaussian as well, but with mean and variance scaled by a factor $m$, and \eqref{eq:colc1}--\eqref{eq:colc3} thus follows by direct application of Theorem \ref{th:stat} and Corollary \ref{col:stat}.
\end{IEEEproof}

\subsection{DLLout metric (Gaussian variance change)}
Now, we analyze the characterization of the LLR for the DLLout metric, which is modeled as \cite{VTC,ION}
\hyp{\label{eq:DLLmodel}}{\Xd}{0}{\sigddo}{0}{\sigddu}{,}
with $\Xd$ the DLLout samples, $\sigddo$ the known variance under $\Ho$, and $\sigddu$ the unknown variance under $\Hu$. Thereby, we model the appearance of multipath as a change in the variance of the DLLout samples, obtaining the following result. 
\begin{col}\label{col:DLL}
Let $\stdd$ and $\sigma^2_{\mathrm{d},1}$ be the tuned change parameter, as in \eqref{eq:PL}, and the actual change parameter, respectively, and let $\sigdo$ be known, then the LLR for the DLL is given by
\eq{\label{eq:LLRd}
\LLR_\mathrm{d}(n) = a\Xd^2 + c,}
with
\eq{
a = \frac{\stdd - \sigddo}{2\sigddo\stdd}; \quad c = \ln\p{\frac{\sigdo}{\std}}.
}
Now, let $\Tdll(\hd)$ be the FMA stopping time for the DLLout metric with LLR in \eqref{eq:LLRd} and threshold $\hd$ so that $\Pfa(\Tdll(\hd),\ma)\leq \at$, and let $\Gamma_m(x)$ denote the cdf of the chi-squared distribution with $m$ degrees of freedom, and $k_i = \sigma^2_{\mathrm{d},i}a$, with $i=\{0,1\}$. Hence,
\begin{align}
\label{eq:cold1}
h_\mathrm{d}\p{\at} &= k_0\Gamma_m^{-1}\p{\p{1-\tilde{\alpha}}^{1/\ma}} + mc,\\
\label{eq:cold2}
\Pmd\p{\Tdll(\hd),m} &\leq \bdll(\hd,m) = \Gamma_m\p{\frac{h_\mathrm{d} - mc}{k_1}}, \\
\label{eq:cold3}
\Pfa\p{\Tdll(\hd),\ma} &\leq 1 - \corch{\Gamma_m\p{\frac{h_\mathrm{d} - mc}{k_0}}}^{\ma}.
\end{align}
\end{col}
\begin{IEEEproof} 
The change in the DLLout metric is modeled as a Gaussian variance change and thus, substituting $\sigma_0 = \sigdo$, $\sigma_1 = \std$ and $\mu_0 = \mu_1 = 0$ into \eqref{eq:LLRgen}, \eqref{eq:LLRd} follows. Thereby, under $\Hu$, and denoting $S_m^{(\mathrm{d})}\doteq\suim\LLR_\mathrm{d}(i)$, we have
\eq{
\begin{aligned}
S_m^{(\mathrm{d})}|\Hu 	&= a\sum_{n=1}^m \sigddu\p{\frac{\Xd}{\sigdu}}^2 + mc\\
							&= k_1\sum_{n=1}^m X_\mathrm{n}^2 + mc,
\end{aligned}
}
with $X_\mathrm{n} \sim \N(0,1)$ a standard Gaussian random variable. A similar result is obtained under $\Ho$, and then we can write $S_m^{(\mathrm{d})}|\mathcal{H}_i = k_i\tilde{X} + mc$, with $i=\{0,1\}$ and $\tilde{X}$ a chi-squared random variable with $m$ degrees of freedom. Hence,
\eq{\label{eq:zd}
\mathcal{H}_i: \frac{S_m^{(\mathrm{d})} - mc}{k_i} \sim \chi^2_m,
}
where $\chi^2_m$ stands for the chi-squared pdf with $m$ degrees of freedom, and \eqref{eq:cold1}--\eqref{eq:cold3} thus follow by direct application of Theorem \ref{th:stat} and Corollary \ref{col:stat}.
\end{IEEEproof}

\subsection{SAM metric (General Gaussian Change)}
Previously, we have analyzed the characterization of the LLR for the particular cases of having a change in either the mean or variance of a Gaussian distribution. In this section, we analyze the characterization of the LLR for the most general case of having a change in both mean and variance of a Gaussian distribution. This is the case of the SAM-based detection, which is presented in \cite{ICL15} and \cite{ION} as
\hyp{}{\Xs}{\muso}{\sigsso}{\musu}{\sigssu}{,}
with $\Xs$ the SAM samples, $\{\muso,\sigsso\}$ the known mean and variance of $\Xs$ under $\Ho$, respectively, and $\{\musu,\sigssu\}$ the unknown change parameters, denoting the mean and variance of $\Xs$ under $\Hu$, respectively. Thereby, we model the appearance of multipath as a change in the mean and variance of the SAM samples, and then, from \eqref{eq:LLRgen} and after some manipulations, we can write the LLR of the SAM metric as
\eq{
\label{eq:LLRs}
\LLR_{\mathrm{s}}(n) = a_\mathrm{s} \Xs^2 + b_\mathrm{s}\Xs + c_\mathrm{s},}
with
\eq{\label{eq:samparam}
\begin{aligned}
a_\mathrm{s} &= \frac{\stss - \sigsso}{2\sigsso\stss};\quad b_\mathrm{s} = \frac{\sigsso\mts - \stss\muso}{\sigsso\stss};\\
c_\mathrm{s} &= \ln\p{\frac{\sigso}{\sts}} + \frac{\stss\muso^2 - \sigsso\mts^2}{2\sigsso\stss},
\end{aligned}
}
where $\mts$ and $\sts$ are the tuned change parameters.

% CN0 + SAM %
\begin{figure*}[htb]
\begin{minipage}[b]{0.5\textwidth}
  \centering
  \centerline{\includegraphics[width=8cm]{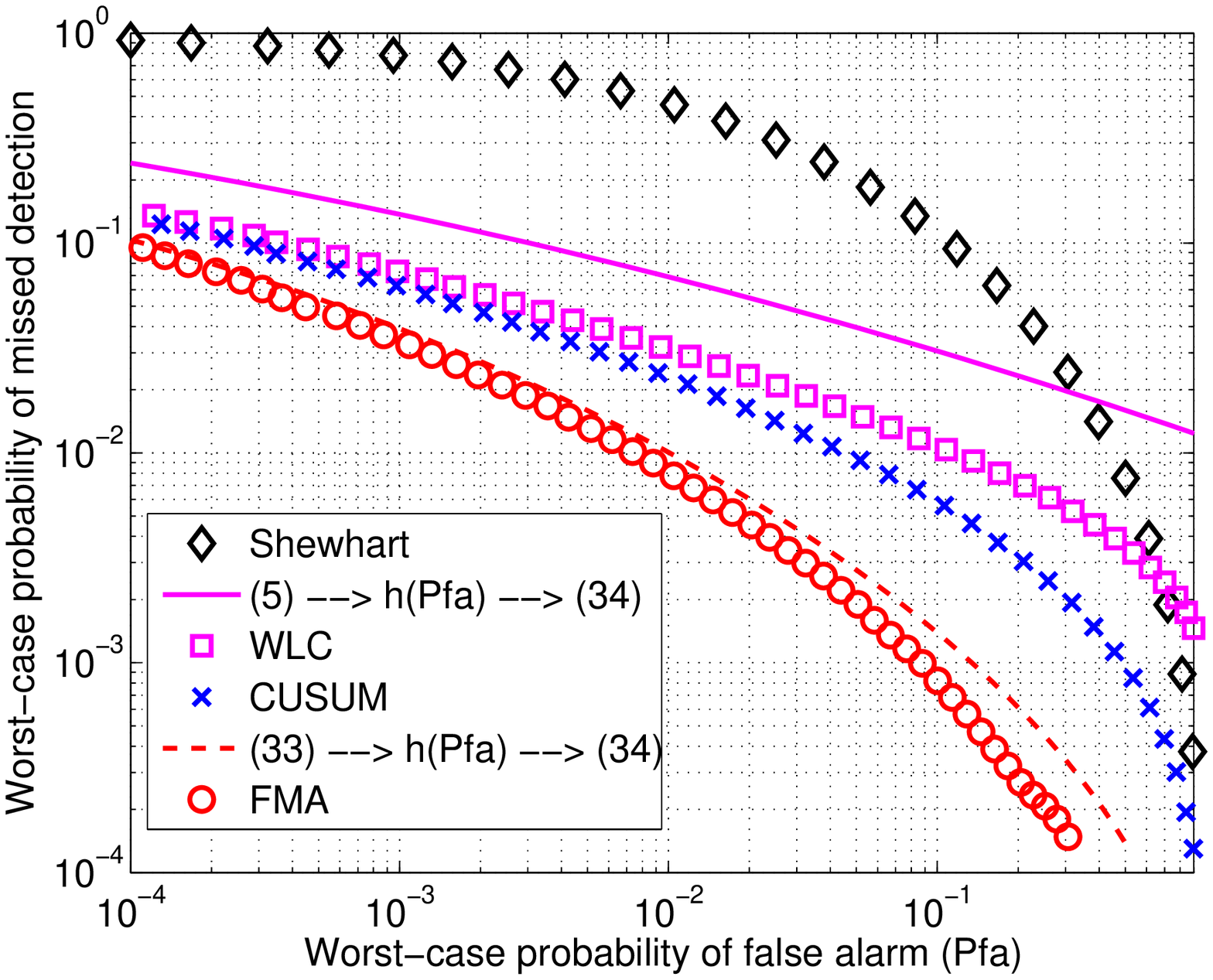}}
\end{minipage}
\hfill
\begin{minipage}[b]{0.5\textwidth}
  \centering
  \centerline{\includegraphics[width=8cm]{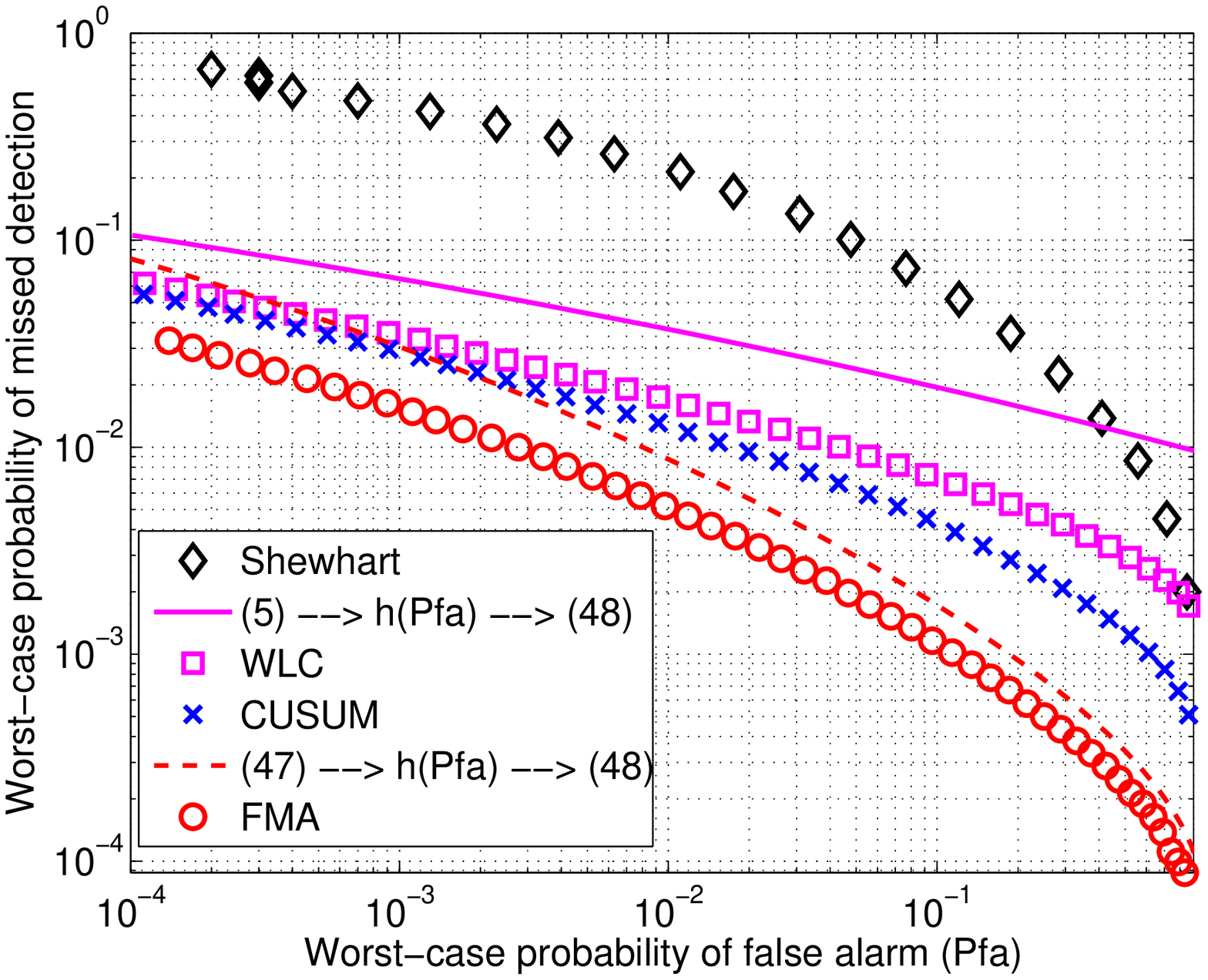}}
\end{minipage}
\caption{Simulated ROC for the \CN (left) and SAM (right) FMA stopping times and their competitors (i.e. CUSUM, WLC and Shewhart), with the theoretical results given in \eqref{eq:colc2}--\eqref{eq:colc3} and \eqref{eq:cols2}--\eqref{eq:cols3}, respectively, and those obtained with \eqref{eq:Tcbound} for the bound of the probability of false alarms.}
\label{fig:ROC1}
\end{figure*}
In this case we cannot find the distribution of $\LLR_{\mathrm{s}}(n)$ in an straightforward way as for the previous cases. Here, in order to find the pdf of $S_m^{(\mathrm{s})} \doteq \sum_{i=1}^m \LLR_{\mathrm{s}}(i)$, we make use of the so-called Edgeworth series approximation and Extreme Value Theory (EVT), which provide a very tight closed-form expression for the bounds for the FMA stopping time of the SAM metric, $\Tsam(\hs)$ \cite{Edgeworth}. For the sake of notational clarity, let us write $S_m^{(\mathrm{s})}$ as the random variable $\Zs$ (i.e. $\Zs = S_m^{(\mathrm{s})}$). Thereby, we can state the following result.
\begin{col}\label{col:sam}
Let $\mu_{\mathrm{s},i}$ and $\sigma^2_{\mathrm{s},i}$, with $i=\{0,1\}$, be the actual mean and variance of the SAM metric under hypothesis $\mathcal{H}_i$, and let $\as$, $\bs$ and $\cs$ be defined as in \eqref{eq:samparam}. Now, let $\Tsam(\hs)$ be the FMA stopping time for the SAM metric with LLR in \eqref{eq:LLRs} and threshold $\hs$ so that $\Pfa(\Tsam(\hs),\ma)\leq\at$, and let $\phi(x)\doteq \N(0,1)$ be the standard normal pdf. Hence, we have
\begin{align}
\label{eq:cols1}
h_\mathrm{s}\p{\at} &= \delta - \frac{\ln\p{-\ln\p{1-\tilde{\alpha}}}}{\gamma},\\
\label{eq:cols2}
\Pmd\p{\Tsam\p{\hs},m} &\leq \bsam(\hs,m) = F_{\mathrm{s},1}\p{\hs}, \\
\label{eq:cols3}
\Pfa\p{\Tsam\p{\hs},\ma} &\leq 1 - \exp\p{-e^{-\gamma(\hs-\delta)}},
\end{align}
with
\eq{\label{eq:Fsam}
\begin{aligned}
\delta 					&= F_{\mathrm{s},0}^{-1}\p{1 - \frac{1}{\ma}},\\
\gamma 					&= \ma f_{\mathrm{s},0}\p{\delta},\\
\end{aligned}
}
and
\eq{\label{eq:Fsam_edg}
\begin{aligned}
F_{\mathrm{s},i}(z) 	&= \Phi(\tilde{z}_i) - \sigma_{\zs,i}\phi(\tilde{z}_i) \sum_{k \in \mathcal{A}} C_{k,\mathcal{H}_i}H_{k-1}(\tilde{z}_i),\\
f_{\mathrm{s},0}(z)	&= \phi(\tilde{z}_0) \corch{1+\sum_{k \in \mathcal{A}} C_{k,\mathcal{H}_0}H_{k}(\tilde{z}_0)},
\end{aligned}
}
where $\mathcal{A} = \{3,4,6\}$, $C_{k,\mathcal{H}_i}$, with $i=\{0,1\}$, are the coefficients $C_k$ (expressions can be found in Appendix \ref{app:colsam}) under $\mathcal{H}_i$, $H_k(z)$ is the Hermite polynomial of degree $k$ evaluated at $z$ and $\tilde{z}_i=(z-\mu_{\zs,i})/\sigma_{\zs,i}$, with
\eq{\label{eq:muz}
\begin{aligned}
\mu_{\zs,i} &= m\corch{\as\p{\sigma^2_{\mathrm{s},i} + \mu^2_{\mathrm{s},i}} + \bs\mu{_{\mathrm{s},i}} + \cs},\\
\sigma^2_{\zs,i} &= m\corch{\sigma^2_{\mathrm{s},i}\corch{2\as\p{\as\sigma^2_{\mathrm{s},i} + 2\as\mu^2_{\mathrm{s},i} + \bs\mu_{\mathrm{s},i}} + \bs^2}}.
\end{aligned}
} 
%Then the mean and variance of $\Zs$ is given by
\end{col}
%\begin{IEEEproof}
%The proof is given in \cite{Edgeworth} and expressions for the coefficients $C_k$ are given in Appendix \ref{app:colsam}.
%\end{IEEEproof}
%\begin{rem}
%\textcolor{red}{In this case, for the calculation of the detection threshold $h_\mathrm{s}$ we have to solve the equation numerically. BLA BLA BLA}
%\end{rem}

%------------------------------------------------------------------------%
%--------------------- SECTION 5 ----------------------------------------%
%------------------------------------------------------------------------%
\section{Numerical Results}
\label{sec:results}
The aim of this section is to firstly compare the proposed FMA stopping time with other approaches in the literature of TCD. This is done in the setting of sig-RAIM by considering the case of multipath detection. Secondly, the availability of the sig-RAIM algorithm is analyzed for the different used integrity signal metrics and compared with the other approaches. It is worth pointing out that in the previous section we have included the unknown change parameters in the theoretical analysis. Here, we will see the effect of the knowledge of these parameters in the integrity risk calculation.

\subsection{Evaluation of the probability minimization criterion}
\label{sec:optimalresults}
Here, we compare the FMA stopping time with those stopping times currently available in the literature of TCD. This comparison is done with both simulated and available theoretical results. The simulated results include the calculation of the worst-case probability of missed detection (i.e. integrity risk) $\Pmd(\Tf(h),m)$ as a function of the simulated worst-case probability of false alarm $\Pf(\Tf(h),\ma)\leq\at$, henceforth referred to as the ROC. The theoretical results include those obtained in Section \ref{sec:metrics} for the FMA stopping time of the different metrics used in this work, stated in Corollaries \ref{col:CN0}, \ref{col:DLL} and \ref{col:sam}. On the other hand, we use the bounds available in the literature for the CUSUM and WLC. For the probability of missed detection of these two methods we have an upper bound in the form of \eqref{eq:beta} (see \cite{Nikiforov} and \cite{Guepie} for the WLC and CUSUM, respectively). For the false alarm probability, we use the upper bound given by \eqref{eq:Pfac}--\eqref{eq:Tcbound}, which holds for both the CUSUM and WLC.

Fig. \ref{fig:ROC1} shows the simulated ROC for the FMA stopping time corresponding to the \CN (left) and SAM (right) metrics. The simulations are obtained with $10^6$ Monte-Carlo runs. The obtained theoretical results in Corollaries \ref{col:CN0} and \ref{col:sam} are also shown, respectively. That is, the bound for $\Pmd$ in \eqref{eq:colc2} and \eqref{eq:cols2}, respectively, as a function of the bound for $\Pfa$ in \eqref{eq:colc3} and \eqref{eq:cols3}, respectively. For the WLC and CUSUM the same expression for $\Pmd$ is used, but taking into account the fixed threshold (as function of $\at$) from \eqref{eq:Tcbound}. This is done by considering the following parameters for the \CN metric, suggested in \cite{ION}: $\muc{_{,0}}=10^{4.4}$, $\mtc=\muc{_{,1}}=10^{3.7}$ and $\sigcc =(\Delta/3)^2$, with $\Delta = \muc{_{,0}}(10^{0.3}-1)$. Moreover, we use $m=6$, $\ma = 60$ assuming a sampling rate of one second for the \CN and $\tta=6$ s and $\ta=1$ min.

On the other hand, for the SAM metric we use the following parameters, similar to those in \cite{ICL15} and \cite{ION}: $\muso=0.1$, $\sigsso = 1.14\cdot 10^{-3}$, $\mts = \musu = 0.2$ and $\stss=\sigssu= 2.03\cdot 10^{-3}$. In this case, we use $m = 6$, $\ma = 300$ assuming a sampling rate of one second for the SAM and $\tta = 6$ s and $\ta = 5$ min. It can be concluded from Fig. \ref{fig:ROC1} that the FMA stopping time outperforms (for both the \CN and SAM metrics), in the sense of the optimality criterion in \eqref{eq:transient1}--\eqref{eq:transient2}, all the other stopping times considered. Moreover, it is worth pointing out that the bounds will also impact the real performance of the methods because the threshold $h$ is in practice fixed using the bounds, with the aim of fixing certain desirable performance. Therefore, the availability of tight bounds is important not only for a theoretical study but also for setting the threshold in practice and provide a level of performance that is close to the desired one. We see how the improvement of the FMA bounds with respect to those available in the literature for other stopping times is quite significant, providing between half- to two-orders-of-magnitude improvement. This is something that, as we will see next, greatly contributes on improving the availability of the sig-RAIM.

%DLL%
\begin{figure}[tb]
  \centering
  \centerline{\includegraphics[width=8cm]{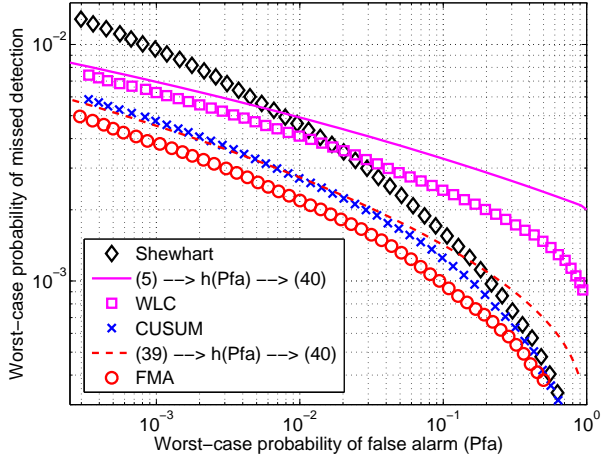}}
\caption{Simulated ROC for the DLLout FMA stopping time and its competitors (i.e. CUSUM, WLC and Shewhart), with the theoretical results given in \eqref{eq:cold2}--\eqref{eq:cold3} and those obtained with \eqref{eq:Tcbound} for the bound of the probability of false alarms.}
\label{fig:ROC2}
\end{figure}
Fig. \ref{fig:ROC2} shows similar results but for the case of the DLLout metric and using the following parameters suggested in \cite{VTC} and \cite{ION}: $\sigddo=(\Delta_0/3)^2$, $\stdd=\sigddu=(\Delta_1/3)^2$, with $\Delta_0 = 0.01$ chips and $\Delta_1 = 0.07$ chips, and $m=6$, $\ma = 60$ assuming a sampling rate of one second for the DLLout and $\tta=6$ s and $\ta=1$ min. The simulations are obtained with $10^{6}$ Monte-Carlo runs. The theoretical results in Corollary \ref{col:DLL} are also shown; that is the bound for $\Pmd$ in \eqref{eq:cold2} as a function of the bound for $\Pfa$ in \eqref{eq:cold3}. This is for the FMA stopping time, for the WLC and CUSUM we use \eqref{eq:cold2} but with the threshold fixed from \eqref{eq:Tcbound}. In this case we also see an improvement of the FMA stopping time with respect to the CUSUM and WLC. Moreover, the improvement in terms of fixed bounds for the FMA with respect to the CUSUM and WLC is also confirmed. Finally, it is worth remarking that the Shewhart stopping time is not giving the best results nether for the DLL nor for the \CN and SAM, making evident the loss of its optimality properties for $m>1$.

\subsection{Numerical example: sig-RAIM availability}
This section is intended to show the behavior of the presented metrics in terms of availability of the sig-RAIM. 

\textbf{CASE 1: \CN metric}\\
Let us start with the \CN metric assuming that we use $m=6$, $\ma = 60$, $\mu_{\mathrm{c},0}=10^{4.4}$ and $\sigcc = 2.5\cdot 10^{4}$, as previously. Also let us suppose we have a tolerable error equivalent to a mean change in the \CN of 7dB; thus we fix the change parameter as $\mtc = 10^{3.7}$, but the actual change parameter is $\mu_{\mathrm{c},1} = 10^{3.4}$. Therefore, for the case of using the FMA stopping time, fixing the detection threshold from \eqref{eq:colc1} to $\hc = 2.92$ so that $\Pfa(\Tcn(\hc),\ma)\leq\at= 10^{-1}$, and substituting the previous values in \eqref{eq:muvarc} and \eqref{eq:colc2}, the integrity risk is bounded as $\Pmd(\Tcn(\hc),m) \leq \bcn(\hc,m) = 6.97\cdot 10^{-4}$. For the CUSUM or WLC stopping time we fix the threshold $\tilde{h}_{\mathrm{c}}$ from \eqref{eq:Tcbound}, which for $\at= 10^{-1}$ gives $\tilde{h}_{\mathrm{c}} = 6.40$, and thus from \eqref{eq:colc2} we get $\Pmd(\Tw(\tilde{h}_{\mathrm{c}}),m) \leq \bcn(\tilde{h}_{\mathrm{c}},m) = 4.56\cdot 10^{-3}$. 

Now, suppose the maximum allowed integrity risk is $\bt = 10^{-2}$; then since $\bcn(\hc) < \bcn(\tilde{h}_{\mathrm{c}}) < \bt$ the sig-RAIM will be available in the case of using any of the analyzed stopping times. On the other hand, we suppose we need $\at = 10^{-2}$ so from \eqref{eq:colc1} and \eqref{eq:Tcbound} we get $\hc = 3.59$ for the FMA and $\tilde{h}_{\mathrm{c}} = 8.70$ for the CUSUM and WLC. Thus, from \eqref{eq:colc2} we have that $\bcn(\hc) = 1.02\cdot 10^{-3}$ and $\bcn(\tilde{h}_{\mathrm{c}}) = 1.33\cdot 10^{-2}$. Hence, in this case, the sig-RAIM will be available only if the FMA stopping time is used. Otherwise, it will not be available since $\bcn(\tilde{h}_{\mathrm{c}})>\bt$. This shows the improvements of the suggested FMA stopping time and the proposed theoretical bounds in terms of the sig-RAIM availability.

\textbf{CASE 2: DLL metric}\\
For the DLL we use $m=6$, $\ma = 60$ and $\sigma^2_{\mathrm{d},0} = 1.11\cdot 10^{-5}$ as in Section \ref{sec:optimalresults}. Now, we evaluate the effect of using the actual change parameter on the availability of the sig-RAIM when a threat is present and we use the tuning parameter $\stdd$. To do so, imagine that a change is present with $\sigma^2_{\mathrm{d},1}=5.44\cdot 10^{-4}$, but the maximum tolerable error in the measured range within the GNSS receiver for each satellite is equal to 14.65 m. For a GPS signal, 14.65 m of error is equivalent to a variation of $\pm 0.05$ chips, which converted to DLL variance as in \cite{ION} gives a minimum detectable change parameter of $\stdd = 2.78\cdot 10^{-4}$. Assuming we want $\Pfa(T(h),\ma)\leq\at = 10^{-2}$, from \eqref{eq:cold1}, we have $\hd = 3.14$ for the FMA and, from \eqref{eq:Tcbound}, $\tilde{h}_{\mathrm{d}}=8.70$ for the CUSUM and WLC. Thereby, if we fix the actual parameter as $\sigma^2_{\mathrm{d},1}=\stdd$, we get from \eqref{eq:cold2} $\bdll(\hd,m) = 1.70\cdot 10^{-2}$ and $\bdll(\tilde{h}_{\mathrm{d}},m) = 4.25\cdot 10^{-2}$, and then, since they are above $\bt$, the sig-RAIM is not available. On the other hand, if we know the actual change parameter, from \eqref{eq:cold2} we have $\bdll(\hd,m) = 2.74\cdot 10^{-3}$ and $\bdll(\tilde{h}_{\mathrm{d}},m) = 7.41\cdot 10^{-3}$, which are bellow $\bt$ and thus the sig-RAIM would be available using either the FMA, CUSUM or WLC stopping time. With this result we corroborate the improvements on the availability by knowing the real change parameter in \eqref{eq:cold2}.

\textbf{CASE 3: SAM metric}\\
Finally, for the SAM metric we use $m=6$, $\ma = 300$, $\muso=0.1$ and $\sigsso = 1.14\cdot 10^{-3}$ as in Section \ref{sec:optimalresults}. For the fixed change parameters, we suppose that the maximum tolerable error leads, from \eqref{eq:PL}, to $\mts = 0.2$ and $\stss= 2.03\cdot 10^{-3}$. Also suppose that $\musu=\mts$ and $\sigssu = \stss$, then if we want $\Pfa(T(h),\ma)\leq\at =  10^{-2}$, from \eqref{eq:cols1}, we have $\hs = 5.53$ for the FMA and, from \eqref{eq:Tcbound}, $\tilde{h}_{\mathrm{s}}=10.31$ for the CUSUM and WLC. Thus, from \eqref{eq:cols2} we get $\Pmd(\Tsam(\hs),m) \leq \bsam(\hs,m) = 8.75\cdot 10^{-3}$ and $\Pmd(\Tsam(\tilde{h}_{\mathrm{s}}),m) \leq \bsam(\tilde{h}_{\mathrm{s}},m) = 3.71\cdot 10^{-2}$. If we wish $\bt = 10^{-2}$, then since $\bsam(\hd,m) \leq \bt$ but $\bsam(\tilde{h}_{\mathrm{s}},m) > \bt$ the sig-RAIM will be available only if the FMA stopping time is used, showing again the improvements in terms of availability.

%------------------------------------------------------------------------%
%--------------------- SECTION 6 ----------------------------------------%
%------------------------------------------------------------------------%
\section{Conclusions}
\label{sec:conclusions}
This work has investigated the problem of \emph{transient} change detection in the context of GNSS signal integrity. We have proposed the use of an FMA stopping time, inspired by the fact that the optimal windowed CUSUM for the case of a Gaussian mean change is the FMA stopping time. This optimality is based on the criterion of minimizing the probability of missed detection with a constraint in the false alarms, which perfectly fits in the GNSS integrity problem. The statistical performance of the FMA stopping time has been theoretically investigated and compared by numerical simulations to different methods available in the literature. This has been done in the setting of GNSS signal integrity by considering the case of multipath detection. These experiments have confirmed the goodness of the presented theoretical results in order to be used for signal integrity. Moreover, it has also been shown that the proposed solution outperforms other solutions available in the literature of \emph{transient} change detection, benefiting not only the performance but the availability of the sig-RAIM algorithm.

%------------------------------------------------------------------------%
%------------------------- APPENDICES -----------------------------------%
%------------------------------------------------------------------------%
\appendices

\section{Proof of Theorem \ref{th:stat}}
\label{app:stat}
The proof of Theorem \ref{th:stat} is divided in three parts. Firstly, we prove the bound for the probability of false alarm given in \eqref{eq:Pfaineq}--\eqref{eq:alpha}. Secondly, we show the proof of the bound for the probability of missed detection given by \eqref{eq:Pmdineq}--\eqref{eq:beta}. Finally, we conclude the proof of the theorem by proving \eqref{eq:h}--\eqref{eq:Pmdh}.

\subsection{Probability of false alarm $\Pfa(\Tf(h),\ma)$}
We first introduce an important result, stated in the following lemma, that will be very useful for proving Theorem \ref{th:stat}.
\begin{lemma}\label{lem:asoc}
Let $S_n\doteq \suim \LLR(i)$, $k\geq m$ and $N > k$ be integers, then
\eq{\label{eq:asoc}
\Pi\p{\bigcap_{i=k}^{k+N-1}\llav{S_i < h}} \geq \corch{\Pi\p{S_m < h}}^N.
}
\end{lemma}
\begin{IEEEproof}
Let $y_i = \LLR(i)$, then from \eqref{eq:Tfma}, for $n\geq m$ we can write
\eq{
S_n = \sum_{i=n-m+1}^n y_i = \sum_{i=1}^n c_{n-i}y_i,
}
with
\eq{
c_i = 
\begin{cases}
1 & \mbox{if } 0\leq i \leq m-1\\
0 & \mbox{if } i\geq m
\end{cases},
}
so that $S_n$ is written as a monotonically increasing function of $\{y_1,\dots,y_n\}$ (since $c_i\geq 0$). Therefore, since $y_1,y_2,\dots$ are iid under $\Pi$, from Theorem 5.1 of \cite{Asoc}, we have that
\eq{
\Pi\p{\inters_{i=k}^{k+N-1}\llav{S_i<h}} \geq \prod_{i=k}^{k+N-1} \Pi\p{S_i < h},
}
and the inequality \eqref{eq:asoc} thus follows from the fact that the distribution of $S_i$, under $\Pi$, is the same for any $i\geq m$.
\end{IEEEproof}

Now, we aim to prove first another useful result for obtaining  \eqref{eq:Pfaineq}--\eqref{eq:alpha}, that is
\eq{\label{eq:Pfaeq}
\Pfa\p{\Tf(h),\ma} = \Pi\p{m\leq \Tf(h) < m+\ma}.
}
To do so, in a similar way as in \cite{Guepie}, from \eqref{eq:transient2} we can write
\eq{
\Pfa(\Tf(h),\ma) = \sup_{l\geq m}\sum_{k=l}^{l+\ma-1} \Pi(\Tf(h) =k).
}
Now denoting $V_l = \Pi(l\leq \Tf(h) < l+\ma)$ for $l\geq m$ and $U_k = \Pi(\Tf(h)=k)$, we have that
\eq{\label{eq:18}
\Pfa(\Tf(h),\ma) = \sup_{l\geq m} V_l = \sup_{l\geq m} \sum_{k=l}^{l+\ma-1}U_k.
}
It is easy to verify, from the definition of $\Tf(h)$ in \eqref{eq:Tfma}, that
\eq{
U_m = \Pi\p{S_m\geq h}
}
and
\eq{
\begin{aligned}
U_{m+1} &= \Pi\p{\llav{S_m<h}\bigcap\llav{S_{m+1}\geq h}} \\
 		&\leq\Pi\p{\llav{S_{m+1}\geq h}} = \Pi\p{\llav{S_{m}\geq h}}=U_m,
\end{aligned}
}
where the second to last equality follows because $S_n$ has the same distribution, under $\Pi$, for $n\geq m$. Similarly, for $k>m$, we have
\eq{
\begin{aligned}
U_{k+1} &= \Pi\p{\inters_{n=m}^{k}\llav{S_n<h} \bigcap \llav{S_{k+1}\geq h}}\\
        &\leq \Pi\p{\inters_{n=m+1}^{k}\llav{S_n<h} \bigcap \llav{S_{k+1}\geq h}} \\
        &= \Pi\p{\inters_{n=m}^{k-1}\llav{S_n<h} \bigcap \llav{S_{k}\geq h}}= U_k.
\end{aligned}
}
Thus, $\{U_k\}_{k\geq m}$ is a non-increasing sequence, and then
\eq{
V_l - V_{l+1} = \sum_{k=l}^{l+\ma-1}U_k - \sum_{k=l+1}^{l+\ma}U_k = U_l - U_{l+\ma} \geq 0,
}
so that $\{V_l\}_{l\geq m}$ is a non-increasing sequence as well. Hence, from \eqref{eq:18} and the definition of $V_l$, 
\eq{\label{eq:22}
\Pfa\p{\Tf(h),\ma} = \sup_{l\geq m} V_l = \Pi\p{m\leq \Tf(h) < m+\ma},
}
and thus \eqref{eq:Pfaeq} follows.

Now, we can proceed with the calculation of $\Pfa(\Tf(h),\ma)$. However, the exact calculation from \eqref{eq:22} is complicated to obtain, and then the calculation of an upper bound is proposed instead. From \eqref{eq:22} and since $\Tf(h)\geq m$ from the definition in \eqref{eq:Tfma}, we can write
\eq{
\Pfa\p{\Tf(h),\ma} = 1 - \Pi\p{\Tf(h) \geq m + \ma},
}
with
\eq{\label{eq:33}
\Pi\p{\Tf(h) \geq m + \ma} = \Pi\p{\inters_{n=m}^{m+\ma-1}\llav{S_n<h}}.
}
So, the proof of \eqref{eq:Pfaineq} and \eqref{eq:alpha} follows by direct application of \eqref{eq:asoc} to \eqref{eq:33}.

\subsection{Probability of missed detection $\Pmd(\Tf(h),m)$}
Applying the Bayes rule in \eqref{eq:transient1} we have
\eq{\label{eq:35}
\begin{aligned}
\Pmd(\Tf(h),m) &= \sup_{v> m} \frac{\Pv\p{\Tf(h)\geq v+m}}{\Pv\p{\Tf(h)\geq v}}\\
          &= \sup_{v> m} \frac{\Pv\p{\inters_{n=m}^{m+v-1}\llav{S_n<h}}}{\Pv\p{\inters_{n=m}^{v-1}\llav{S_n<h}}},
\end{aligned}}
where the last equality follows from the definition of $\Tf(h)$ in \eqref{eq:Tfma}. Due to the windowed behavior of $\Tf(h)$ we have assumed that $v> m$. As for $\Pfa(\Tf(h),\ma)$, the exact calculation of $\Pmd(\Tf(h),m)$ from \eqref{eq:35} is quite difficult, and then we propose the derivation of an upper bound.

Now, letting the event $\mathcal{A}_n = \{S_n<h\}$, with $n\geq m$, it is clear that $\mathcal{A}_{v-1}$ and $\mathcal{A}_{m+v-1}$ are independent because they do not share any samples, thus
\eq{
\Pv\p{\inters_{n=m}^{m+v-1}\mathcal{A}_n} \leq \Pv\p{\mathcal{A}_{m+v-1}}\Pv\p{\inters_{n=m}^{v-1}\mathcal{A}_n},
}
since in the left side we evaluate more events than in the right side. So, applying this result to \eqref{eq:35} we have that
\eq{
\begin{aligned}
\Pmd\p{\Tf(h),m} &\leq \sup_{v> m} \Pv\p{S_{m+v-1}<h} \\
            &= \Po\p{S_m < h} ,
\end{aligned}
}
where the equality follows because $S_{m+v-1}$ is identically distributed under $\Pv$, and \eqref{eq:Pmdineq}--\eqref{eq:beta} thus follow, completing the proof of Theorem \ref{th:stat}.

\section{Calculation of coefficients $C_k$}
\label{app:colsam}
A complete proof of the results in Corollary \ref{col:sam} can be found in \cite{Edgeworth}. Here we only give the expression for $C_k$ so that the bounds in Corollary \ref{col:sam} can be calculated. The coefficients $C_{k,\mathcal{H}_i}$ are obtained as
\eq{\label{eq:coef}
\begin{aligned}
C_3 &= \frac{\xi_{\zs,3} -3\mu_{\zs,i}\xi_{\zs,2} + 2\mu_{\zs,i}^3}{\sigma^3_{\zs,i}},\\
C_4 &= \frac{\xi_{\zs,4} - 4\mu_{\zs,i}\xi_{\zs,3} + 6\mu_{\zs,i}^2\xi_{\zs,2} -3\mu_{\zs,i}^4}{\sigma^4_{\zs,i}} - 3,\\
C_6 &= 10C_3^2,
\end{aligned}
}
with $\xi_{\zs,n}$ the moments of $\Zs$, under $\mathcal{H}_i$, given by 
\eq{\label{eq:mzs}
\begin{aligned}
\xi_{\zs,2} &= \mu_{\zs,i}^2 + \sigma_{\zs,i}^2,\\
\xi_{\zs,3} &= m\corch{\xi_{\ys,3} + (m-1)\p{3\mu_{\zs,i}\xi_{\ys,2} + (m-2)\mu_{\zs,i}^3}},\\
\xi_{\zs,4} &= m\corch{\xi_{\ys,4} + (m-1)\p{\Omega + (m-2)\p{\Gamma + \Lambda}}},
\end{aligned}
}
where $\mu_{\zs,i}$ and $\sigma_{\zs,i}^2$ are given by \eqref{eq:muz}, $\Omega = 4\mu_{\zs,i}\xi_{\ys,3}+3\xi_{\ys,2}^2$, $\Gamma = 6\mu_{\zs,i}^2\xi_{\ys,2}$, $\Lambda = (m-3)\mu_{\zs,i}^4$, and $\xi_{\ys,k}$ are the moments of $Y_n$ given by
\eq{\label{eq:mys}
\xi_{\ys,n} = \sum_{i=0}^n A(i),
}
with
\eq{\label{eq:Ai}
A(i) = \sum_{j=0}^i \binom{n}{i}\binom{i}{j} \as^{n-i} \bs^{i-j} \cs^j \xi_{\xs,2n-i-j},
}
where $\binom{n}{i} = (n!)/(i!(n-i)!)$ is the binomial coefficient, $\xi_{\xs,k}\doteq \E_i(\Xs^k)$ is the moment of $k$-th order of $\Xs$ under hypothesis $\mathcal{H}_i$ and \{$\as$, $\bs$, $\cs$\} are defined in \eqref{eq:samparam}.

\bibliographystyle{IEEEtran}
\bibliography{biblio}

\begin{thebibliography}{10}
\providecommand{\url}[1]{#1}
\csname url@samestyle\endcsname
\providecommand{\newblock}{\relax}
\providecommand{\bibinfo}[2]{#2}
\providecommand{\BIBentrySTDinterwordspacing}{\spaceskip=0pt\relax}
\providecommand{\BIBentryALTinterwordstretchfactor}{4}
\providecommand{\BIBentryALTinterwordspacing}{\spaceskip=\fontdimen2\font plus
\BIBentryALTinterwordstretchfactor\fontdimen3\font minus
  \fontdimen4\font\relax}
\providecommand{\BIBforeignlanguage}[2]{{%
\expandafter\ifx\csname l@#1\endcsname\relax
\typeout{** WARNING: IEEEtran.bst: No hyphenation pattern has been}%
\typeout{** loaded for the language `#1'. Using the pattern for}%
\typeout{** the default language instead.}%
\else
\language=\csname l@#1\endcsname
\fi
#2}}
\providecommand{\BIBdecl}{\relax}
\BIBdecl

\bibitem{salcedo}
G.~Seco-Granados \emph{et~al.}, ``{Challenges in indoor global navigation
  satellite systems: unveiling its core features in signal processing},''
  \emph{IEEE Signal Processing Magazine}, vol.~29, no.~2, pp. 108--131, 2012.

\bibitem{SBAS}
P.~W. Bradford, J.~Spilker, and P.~Enge, \emph{{Global Positioning System:
  Theory and Applications}}.\hskip 1em plus 0.5em minus 0.4em\relax Aiaa, 1996,
  vol.~2.

\bibitem{RTCA}
R.~SC-159, ``{Minimum operational performance standards for global positioning
  system/wide area augmentation system airborne equipment},'' \emph{RTCA Inc.
  Document No. RTCA/DO-229D}, 2006.

\bibitem{RAIM1}
B.~W. Parkinson and P.~Axelrad, ``{Autonomous GPS integrity monitoring using
  the pseudorange residual},'' \emph{Navigation}, vol.~35, no.~2, pp. 255--274,
  1988.

\bibitem{RAIM2}
R.~G. Brown and P.~McBurney, ``{Self-contained GPS integrity check using
  maximum solution separation},'' \emph{Navigation}, vol.~35, no.~1, pp.
  41--53, 1988.

\bibitem{integrity}
D.~Salos \emph{et~al.}, ``{Receiver autonomous integrity monitoring of GNSS
  signals for electronic toll collection},'' \emph{IEEE Transactions on
  Intelligent Transportation Systems}, vol.~15, no.~1, pp. 94--103, 2014.

\bibitem{GMV}
J.~Cosmen-Schortmann \emph{et~al.}, ``Integrity in urban and road environments
  and its use in liability critical applications,'' in \emph{Proc. IEEE/ION
  Position, Location and Navigation Symposium}, 2008, pp. 972--983.

\bibitem{GMV2}
R.~Toledo-Moreo \emph{et~al.}, ``An analysis of positioning and map-matching
  issues for {GNSS}-based road user charging,'' in \emph{Proc. IEEE Intelligent
  Transportation Systems (ITSC)}, 2010, pp. 1486--1491.

\bibitem{Poor}
H.~V. Poor and O.~Hadjiliadis, \emph{{Quickest Detection}}.\hskip 1em plus
  0.5em minus 0.4em\relax Cambridge University Press, 2009.

\bibitem{Baseville}
M.~Basseville and I.~V. Nikiforov, \emph{{Detection of Abrupt Changes: Theory
  and Application}}.\hskip 1em plus 0.5em minus 0.4em\relax Englewood Cliffs:
  Prentice Hall, 1993.

\bibitem{Page}
E.~Page, ``Continuous inspection schemes,'' \emph{Biometrika}, pp. 100--115,
  1954.

\bibitem{Lorden}
G.~Lorden, ``{Procedures for reacting to a change in distribution},'' \emph{The
  Annals of Mathematical Statistics}, vol.~42, no.~6, pp. 1897--1908, 1971.

\bibitem{Moustakides1}
G.~V. Moustakides, ``Optimal stopping times for detecting changes in
  distributions,'' \emph{The Annals of Statistics}, pp. 1379--1387, 1986.

\bibitem{Bojdecki}
T.~Bojdecki, ``{Probability maximizing approach to optimal stopping and its
  application to a disorder problem},'' \emph{Stochastics}, vol.~3, pp. 61--71,
  1979.

\bibitem{Pollak}
M.~Pollak and A.~M. Krieger, ``{Shewhart revisited},'' \emph{Sequential
  Analysis}, vol.~32, pp. 230--242, 2013.

\bibitem{Moustakides2}
G.~V. Moustakides, ``{Multiple optimality properties of the Shewhart test},''
  \emph{Sequential Analysis}, vol.~33, pp. 318--344, 2014.

\bibitem{interf1}
W.~Sun and M.~G. Amin, ``A self-coherence anti-jamming {GPS} receiver,''
  \emph{IEEE Transactions on Signal Processing}, vol.~53, no.~10, pp.
  3910--3915, 2005.

\bibitem{interf2}
Y.~D. Zhang and M.~G. Amin, ``Anti-jamming {GPS} receiver with reduced phase
  distortions,'' \emph{IEEE Signal Processing Letters}, vol.~19, no.~10, pp.
  635--638, 2012.

\bibitem{MP1}
H.~K. Lee, J.~G. Lee, and G.-I. Jee, ``{GPS} multipath detection based on
  sequence of successive-time double-differences,'' \emph{IEEE Signal
  Processing Letters}, vol.~11, no.~3, pp. 316--319, 2004.

\bibitem{MP2}
A.~Giremus, J.-Y. Tourneret, and V.~Calmettes, ``A particle filtering approach
  for joint detection/estimation of multipath effects on {GPS} measurements,''
  \emph{IEEE Transactions on Signal Processing}, vol.~55, no.~4, pp.
  1275--1285, 2007.

\bibitem{spoof1}
M.~L. Psiaki and T.~E. Humphreys, ``{GNSS} spoofing and detection,''
  \emph{Proceedings of the IEEE}, vol. 104, no.~6, pp. 1258--1270, 2016.

\bibitem{spoof2}
E.~Axell, E.~G. Larsson, and D.~Persson, ``{GNSS} spoofing detection using
  multiple mobile {COTS} receivers,'' in \emph{IEEE International Conference on
  Acoustics, Speech and Signal Processing (ICASSP)}, 2015, pp. 3192--3196.

\bibitem{WSN}
J.-F. Chamberland and V.~V. Veeravalli, ``Wireless sensors in distributed
  detection applications,'' \emph{IEEE Signal Processing Magazine}, vol.~24,
  no.~3, pp. 16--25, 2007.

\bibitem{signal_detection}
T.~Osklper and H.~V. Poor, ``{Quickest detection of a random signal in
  background noise using a sensor array},'' \emph{EURASIP Journal on Applied
  Signal Processing}, no.~1, 2005.

\bibitem{cognitive}
E.~Axell \emph{et~al.}, ``Spectrum sensing for cognitive radio:
  State-of-the-art and recent advances,'' \emph{IEEE Signal Processing
  Magazine}, vol.~29, no.~3, pp. 101--116, 2012.

\bibitem{SAM}
D.~Egea, J.~A. L\'{o}pez-Salcedo, and G.~Seco-Granados, ``{Interference and
  multipath sequential tests for signal integrity in multi-antenna GNSS
  receivers},'' in \emph{Proc. IEEE 8th Sensor Array and Multichanel Signal
  Processing Workshop (SAM)}, 2014, pp. 117--120.

\bibitem{ICL14}
D.~Egea, G.~Seco-Granados, and J.~A. L\'{o}pez-Salcedo, ``{Single- and
  multi-correlator sequential tests for signal integrity in multi-antenna GNSS
  receivers},'' in \emph{Proc. International Conference on Localization and
  GNSS (ICL-GNSS)}, 2014, pp. 117--120.

\bibitem{ICL15}
D.~Egea-Roca, G.~Seco-Granados, and J.~A. L\'{o}pez-Salcedo, ``{On the use of
  quickest detection theory for signal integrity monitoring in single-antenna
  GNSS receivers},'' in \emph{Proc. International Conference on Localization
  and GNSS (ICL-GNSS)}, 2015, pp. 1--6.

\bibitem{VTC}
------, ``{Quickest detection framework for signal integrity monitoring in
  low-cost GNSS receivers},'' in \emph{Proc. IEEE 82nd Vehicular Technology
  Conference (VTC Fall)}, 2015, pp. 1--5.

\bibitem{IONint}
------, ``Signal-level integrity and metrics based on the application of
  quickest detection theory to interference detection,'' in \emph{Proc. 28th
  International Technical Meeting of The Satellite Division of the Institute of
  Navigation (ION GNSS+)}, 2015, pp. 3136 -- 3147.

\bibitem{ION}
------, ``Signal-level integrity and metrics based on the application of
  quickest detection theory to multipath detection,'' in \emph{Proc. 28th
  International Technical Meeting of The Satellite Division of the Institute of
  Navigation (ION GNSS+)}, 2015, pp. 2926--2938.

\bibitem{Willett}
C.~Han, P.~K. Willett, and D.~Abraham, ``Some methods to evaluate the
  performance of {P}ages' test as used to detect transient signals,''
  \emph{IEEE Transactions on Signal Processing}, vol.~47, no.~8, pp.
  2112--2127, 1999.

\bibitem{Willett1}
Z.~Wang and P.~K. Willett, ``All-purpose and plug-in power-law detectors for
  transient signals,'' \emph{IEEE Transactions on Signal Processing}, vol.~49,
  no.~11, pp. 2454--2466, 2001.

\bibitem{Willett2}
------, ``{A variable threshold Page procedure for detection of transient
  signals},'' \emph{IEEE Transactions on Signal Processing}, vol.~53, no.~11,
  pp. 4397--4402, 2005.

\bibitem{Nikiforov}
B.~Bakhache and I.~Nikiforov, ``{Reliable detection of faults in measurement
  systems},'' \emph{International Journal of Adaptive Control and Signal
  Processing}, vol.~14, no.~7, pp. 683--700, 2000.

\bibitem{Guepie}
B.~K. Gu{\'e}pi{\'e}, L.~Fillatre, and I.~Nikiforov, ``Sequential detection of
  transient changes,'' \emph{Sequential Analysis}, vol.~31, no.~4, pp.
  528--547, 2012.

\bibitem{SSP}
D.~Egea-Roca \emph{et~al.}, ``{A finite moving average test for transient
  change detection in GNSS signal strength monitoring},'' in \emph{Proc. IEEE
  Statistical Signal Processing Workshop (SSP)}, 2016.

\bibitem{Lai}
T.~L. Lai, ``{Information bounds and quick detection of parameter changes in
  stochastic systems},'' \emph{IEEE Transactions on Information Theory},
  vol.~44, no.~7, pp. 2917--2929, 1998.

\bibitem{Edgeworth}
D.~Egea-Roca, G.~Seco-Granados, and J.~A. L{\'o}pez-Salcedo, ``Closed-form
  approximations for the performance upper bound of inhomogeneous quadratic
  tests,'' \emph{arXiv preprint arXiv:1608.00615}, 2016.

\bibitem{Asoc}
J.~D. Esary \emph{et~al.}, ``Association of random variables, with
  applications,'' \emph{The Annals of Mathematical Statistics}, vol.~38, no.~5,
  pp. 1466--1474, 1967.

\end{thebibliography}
\end{document}